\documentclass[preprint,amsmath,amssymb]{revtex4}
\usepackage{graphicx}
\usepackage{epsfig}
\usepackage{dcolumn}
\usepackage{bm}
\begin{document}
\title{Collective oscillations of a Bose-Fermi mixture: Effect of unequal mass of Bose and Fermi particles}
\author{Arup Banerjee \\
Laser Physics Application Division, Raja Ramanna Centre for Advanced Technology\\
Indore 452013, India\\}
\begin{abstract}
We investigate the effect of different mass of a Bose- and a Fermi-particle on the collective oscillations of the degenerate boson-fermion mixtures. In particular we consider the monopole and the quadrupole modes of the oscillations and study their characters and the frequencies by using variational-sum-rule approach. We find that for both the modes there exists a critical value of the ratio of boson-fermion mass below and above which the character and the frequency of the respective modes are significantly different. 
\end{abstract}

 \maketitle 
\section{Introduction}
In recent years many theoretical and experimental studies involving dilute mixture of trapped ultra cold gases of bosonic and fermionic atoms have been reported in the literature. These mixtures provide a convenient way to achieve degenerate fermionic gas by means of sympathetic cooling as the conventional evaporative cooling methods used to obtain Bose-Einstein condensates (BEC) are not applicable to fermions. Besides this Bose-Fermi mixtures can be used to study many aspects of quantum statistics.  Experimentally, stable BEC immersed in a degenerate Fermi gas have been realized with $^{7}Li$ in $^{6}Li$ \cite{schreck}, $^{23}Na$ in $^{6}Li$ \cite{hadzibabic}, and $^{87}Rb$ in $^{40}K$ \cite{roati,ospeklaus,zacanti}. The first two Bose-Fermi mixtures are characterized by positive inter-species scattering lengths or repulsive interaction between bosons and fermions. On the other hand, mixture of $^{87}Rb$ and  $^{40}K$ give rise to large attractive boson-fermion interaction. Recently, by exploiting Fesbach resonance the inter-species interaction has been tuned with a large range positive and negative values. 

Theoretically, the trapped system of Bose-Fermi mixture have been studied within the mean-field approximation to determine the bosons and fermion density profiles at zero temperature. These studies mainly focused on the phenomena of phase separation for repulsive and collapse mixture for attractive boson-fermion interaction respectively have been studied extensively \cite{molmer1,molmer2,yabu1,roth,mmodugno,chui,tosi,jezek,adhikari}. Beside ground state density profiles the dynamic properties of Bose-Fermi mixture like free expansion \cite{hu} and the spectrum of collective excitations \cite{yabu2,yabu3,tosi1,liu,tosi2} have also been studied. All these studies have clearly showed that the spectra of collective excitations bear  unambiguous signatures of phase transitions that Bose-Fermi mixtures undergo under appropriate conditions. Moreover, for BEC it is well established that the frequencies of collective oscillations can be measured with very high precision. Consequently, it is natural to expect that by measuring the frequencies of the collective oscillations of a trapped Bose-Fermi mixture it would be possible to determine the parameters at which mixture undergo phase separation or collapse. 

Keeping this in mind we focus our attention on the effect of unequal mass of a Bose- ($m_{b}$) and a Fermi-particle ($m_{f}$) on the frequencies of collective oscillations. At this point we note that most of the studies on the collective oscillations assumed $m_{b}= m_{f}$. In reality, for all the experimental realizations as mentioned above, $m_{b}\neq m_{f}$. For example, in the case of extensively studied Rb-K mixture, $m_{b}/m_{f} \approx 87/40$. In this paper we show that $m_{b}\neq m_{f}$ leads to significant alteration of the nature and the frequencies of the modes of the collective oscillations over the equal mass case \cite{yabu2,liu}. In order to calculate the frequencies of collective oscillations we adopt well established sum rule approach of many-body response theory \cite{bohigas,lipparini}. It has already been shown that the sum rule approach yields quite accurate results for the frequencies of collective oscillations of trapped Bose-Fermi mixtures \cite{yabu3}. Using sum rule approach, we derive  analytical expressions for the frequencies of monopole and quadrupole modes of oscillations of a Bose-Fermi mixture with mass of a Bose-particle being different from that of a Fermi-particle. Our expressions are generalizations of the results of Ref. \cite{yabu2}. These analytical expressions for frequencies are then used to determine critical mass ratio $(m_{b}/m_{f})_{c})$ at which character of the modes undergoes a change resulting in the modifications of the frequencies significantly. 

The content of the paper is as follows. In the next section we present the derivations of analytical expressions for the frequencies followed by a discussion of our results in Section III. The paper is concluded in Section IV.
 
\section{Variational sum rule approach}
The basic result of the sum-rule approach \cite{bohigas,lipparini} is that the upper bound of the
lowest excitation energy is given by
\begin{equation}
\hbar\Omega_{ex} = \sqrt{\frac{m_{3}}{m_{1}}},
\label{frequency}
\end{equation}
where
\begin{equation}
m_{p} = \sum_{n}|\langle 0|F|n\rangle |^{2}\left (\hbar\omega_{n0}\right )^{p},
\end{equation}
is the $p$-th order moment of the excitation energy associated with the
excitation operator $F$ and $\Omega_{ex}$ is the frequency of excitation.
Here $\hbar\omega_{n0}=E_{n}-E_{0}$ is
the excitation energy of eigenstate $|n\rangle$ of the Hamiltonian
$H$.  Moreover, Eq. (\ref{frequency}) can be used for computation of
the excitation energies by exploiting the fact that the moments
$m_{1}$ and $m_{3}$ can be expressed as expectation values of
the commutators between $F$ and $H$ in the ground state
$|0\rangle$. These relations are
\begin{eqnarray}
m_{1} & = & \frac{1}{2}\langle 0|\left [F^{\dagger},\left [H,F\right ]\right ]|0\rangle , \nonumber \\
m_{3} & = & \frac{1}{2}\langle 0|\left [\left [F^{\dagger},H\right ],\left [\left [H,[H,F\right ]\right ]\right ]
|0\rangle .
\label{moments}
\end{eqnarray}
The main advantage of the sum-rule approach is that it allows us
to calculate the dynamic properties like excitation frequencies of
many-body systems with the knowledge of ground state $|0\rangle$
(or the ground-state density) only.  

For the purpose of calculation one needs to choose an appropriate
excitation operator $F$. In this paper we
consider the monopole and the quadrupole modes of the collective oscillations. The excitation operators corresponding to these modes are defined by 
\begin{equation}
F_{\alpha}^{\pm} = f_{\alpha}^{b}({\bf r}) \pm f_{\alpha}^{f}({\bf r}) ,
\label{excitoperators}
\end{equation}
where $\alpha$ stands for monopole or quadrupole modes, the indices $b/f$ denote bosons/fermions  and the functions $f_{\alpha}$ are given by
$f_{M} = r^{2}$ for monople and $f_{Q} = 3z^{2} - r^{2}$ for quadrupole modes. Following \cite{yabu2} we take the excitation operators as a linear combination of $F_{\alpha}^{+}$ and $F_{\alpha}^{-}$  to simulate the effect of mixing of boson and fermion oscillations. It is written as
\begin{equation}
F_{\alpha}(\theta) = F_{\alpha}^{+}cos\theta + F_{\alpha}^{-}sin\theta
\label{lowexcitation}
\end{equation}
where the mixing angle $\theta$ lies between $-\pi/2$ and $\pi/2$. The character of the mode is given by the value of $\theta$, for example, $\theta = \pi/4$ for the bosonic and $-\pi/4$ for the fermionic modes, $\theta = 0$ for the in-phase oscillation, and $\theta = \pi/2$ for the out-of-phase oscillations of the two types of particles. The value of mixing angle for each mode is determined by minimizing the corresponding frequency. The presence of two kinds of particle leads to two types of collective oscillations for each multipole which are orthogonal to each other. Out of these two modes the low-lying mode is excited by the operator of Eq. (\ref{lowexcitation}) and the operator $F_{\alpha}^{+}sin\theta_{l} - F_{\alpha}^{+}cos\theta_{l}$ for each $\alpha$ excites the orthogonal high-lying mode. To study the effect of unequal mass we focus our attention on the low-lying mode only. 

In order to calculate the moments $m_{1}$ and $m_{3}$, we use the expression for the energy functional (in terms of boson $\rho_{b}$ and fermion $\rho_{f}$ densities) of a harmonically trapped Bose-Fermi mixture as given by \cite{albus}

\begin{eqnarray}
E[\rho_{b},\rho_{f}] & = & T_{b}[\rho_{b}] + V_{b}[\rho_{b}] + E_{int}^{bb}[\rho_{b}] \nonumber \\
                     & + & T_{f}[\rho_{f}] + V_{f}[\rho_{f}] + E_{int}^{bf}[\rho_{b},\rho_{f}]\label{energyfunc}
                    \end{eqnarray}
where $T_{b}[\rho_{b}]$ ($V_{b}[\rho_{b}]$) and $T_{f}[\rho_{f}]$ ($V_{f}[\rho_{f}]$) represent the kinetic energy (the trapping energy) functionals of bosons and fermions respectively. The remaining two terms $E_{int}^{bb}[\rho_{b}]$ and $E_{int}^{bf}[\rho_{b},\rho_{f}]$ represent the boson-boson and the boson-fermion interaction energy functionals respectively. In writing above energy functional we have neglected the fermion-fermion interaction energy term as fermions are assumed to be spin polarized. The kinetic energy functional $T_{b}[\rho_{b}]$ is given by
\begin{equation}
T_{b}[\rho_{b}] = \frac{h^{2}}{2m_{b}}\int|\vec{\nabla}\sqrt{\rho_{b}}|d^{3}{\bf r}.
\label{keb}
\end{equation}
On the other hand, exact form of the kinetic energy functional for fermions is not known. In this paper we work within the Thomas-Fermi approximation and use following expression for the kinetic energy functional
\begin{equation}
T_{f}[\rho_{f}] = \frac{h^{2}}{2m_{f}}\frac{3}{5}\left (6\pi^{2}\right )^{2/3}\int\rho_{f}^{5/3}({\bf r})d^{3}{\bf r}.
\label{kef}
\end{equation}
This approximation is quite accurate when the number of fermions is fairly large and in this paper we choose number of fermions ($N_{f}$) such that the TF approximation remains valid.  The harmonic trapping energy functional for spherically symmetric trapping potential are given by
\begin{eqnarray}
V_{b}[\rho_{b}] & = & \frac{1}{2}m_{b}\omega_{b}^{2}\int r^{2}\rho_{b}({\bf r})d^{3}{\bf r} \nonumber \\
V_{f}[\rho_{f}] & = & \frac{1}{2}m_{f}\omega_{f}^{2}\int r^{2}\rho_{f}({\bf r})d^{3}{\bf r},
\label{trapenergy}
\end{eqnarray}
where $\omega_{b}$ and $\omega_{f}$ are frequencies of the trapping potential for bosons and fermions respectively. In accordance with the experimental settings we use $\omega_{f} = \sqrt{m_{b}/m_{f}}\omega_{b}$ for the purpose of our calculations. Finally the forms for the interaction energies within the mean-field approximation are given by
\begin{eqnarray}
E_{int}^{bb}[\rho_{b}] & = & g_{bb}\int \rho_{b}^{2}({\bf r})d^{3}{\bf r} \nonumber \\
E_{int}^{bf}[\rho_{b},\rho_{f}] & = & g_{bf}\int \rho_{b}({\bf r})\rho_{f}({\bf r})d^{3}{\bf r}
\label{intenergies}
\end{eqnarray}
The boson-boson coupling strength is given by $g_{bb} = 4\pi\hbar^{2}a_{bb}/m_{b}$, where $a_{bb}$ is the boson-boson s-wave scattering length. The boson-fermion coupling strength reads $g_{bf} = 4\pi\hbar^{2}a_{bf}/m_{r}$, where $a_{bf}$ is the boson-fermion s-wave scattering length and $m_{r} = m_{b}m_{f}/(m_{b} + m_{f})$ is the reduced boson-fermion mass.

Before proceeding further we wish to note here that representation of ground-state energy as functional of density is ensured by Hohenberg-Kohn theorem of density functional theory \cite{hk}. The ground-state densities $\rho_{b}$ and $\rho_{f}$ can be determined by imposing stationary conditions:
\begin{equation}
\frac{\delta E[\rho_{b},\rho_{f}]}{\delta \rho_{b}} = \mu_{b} ;\qquad \frac{\delta E[\rho_{b},\rho_{f}]}{\delta \rho_{f}} = \mu_{f},
\end{equation}
where $\mu_{b}$ and $\mu_{f}$ are boson and fermion chemical potentials, respectively. The chemical potential are fixed by the normalization conditions $\int\rho_{b}({\bf r})d^{3}{\bf r} = N_{b}$ and $\int\rho_{f}({\bf r})d^{3}{\bf r} = N_{f}$, where $N_{b}$ and $N_{f}$ representing number of bosons and fermions, respectively.
In this paper we exploit this variational nature of the ground-state energy to determine the ground-state density profiles of bosons and fermions. More about this approach will be discussed later in this section. 

Using the above energy functional (Eq. (\ref{energyfunc})) along with the Eq. (\ref{moments}) we find after a tedious although straightforward algebra following expression for the frequency of the monopole mode of collective oscillations
\begin{equation}
\frac{\omega_{m}^{2}}{\omega_{b}^{2}} = \frac{1}{2}\left [ \frac{\left (X_{m} + Y_{m} + Z_{m}\right ) + 2\sin\theta \cos\theta\left (X_{m} - Y_{m}\right ) - 2\sin^{2}\theta Z_{m}}{\left (A_{m} + B_{m}\right ) + 2\sin\theta cos\theta\left (A_{m} - B_{m}\right )}\right ]
\label{monopole}
\end{equation}
with
\begin{eqnarray}
X_{m} & = & 4T_{b} + 4V_{b} + 9E_{int}^{bb} + 9E_{int}^{bf}  + 3\Delta_{f} - \Delta \nonumber \\
Y_{m} & = & \left (\frac{m_{b}}{m_{f}}\right )^{2}\left (4T_{f} + 4V_{f} + 9E_{int}^{bf} + 3\Delta_{b} - \Delta \right )\nonumber \\
Z_{m} & = & 2\left (\frac{m_{b}}{m_{f}}\right )\Delta  \nonumber \\
A_{m} & = & V_{b} \nonumber \\
B_{m} & = & \left (\frac{m_{b}}{m_{f}}\right )V_{f}.
\end{eqnarray}
In the above expressions $T_{b,f}$, $V_{b,f}$, $E_{int}^{bb}$ and $E_{int}^{bf}$ are the respective integrals defined in Eqs. (\ref{keb}) - (\ref{intenergies}) evaluated with the ground-state boson and fermion densities. Besides these energy integrals, remaining three quantitities $\Delta_{f}$, $\Delta_{b}$, and $\Delta$ are given by
\begin{eqnarray}
\Delta_{b} & = &  g_{bf}\int r\rho_{b}({\bf r})\frac{d\rho_{f}({\bf r})}{dr}d^{3}{\bf r} \nonumber \\
\Delta_{f} & = &  g_{bf}\int r\rho_{f}({\bf r})\frac{d\rho_{b}({\bf r})}{dr}d^{3}{\bf r} \nonumber \\
\Delta & = &  g_{bf}\int r\frac{d\rho_{b}({\bf r})}{dr}\frac{d\rho_{f}({\bf r})}{dr}d^{3}{\bf r}. 
\label{delta}
\end{eqnarray}

The expression for the frequency of the quadrupole mode is given by
\begin{equation}
\frac{\omega_{q}^{2}}{\omega_{b}^{2}} = \frac{1}{4}\left [ \frac{\left (X_{q} + Y_{q} + Z_{q}\right ) + 2sin\theta cos\theta\left (X_{q} - Y_{q}\right ) - 2sin^{2}\theta Z_{q}}{\left (A_{q} + B_{q}\right ) + 2sin\theta cos\theta\left (A_{q} - B_{q}\right )}\right ]
\label{quadrupole}
\end{equation}
with
\begin{eqnarray}
X_{q} & = & 8T_{b} + 8V_{b}  -  \frac{4}{5}\Delta \nonumber \\
Y_{q} & = & \left (\frac{m_{b}}{m_{f}}\right )^{2}\left (8T_{f} + 8V_{f} -  \frac{4}{5}\Delta  \right )\nonumber \\
Z_{q} & = & 2\left (\frac{m_{b}}{m_{f}}\right )\frac{4}{5}\Delta 
\label{xqyqzq}
\end{eqnarray}
and $A_{q} = A_{m}, B_{q} = B_{m}$. 

The expressions given by Eqs. (\ref{monopole})- (\ref{quadrupole}) constitute the main results of this paper and they correctly reduce to the corresponding equations of Ref. \cite{yabu2} for the case of $m_{b}/m_{f} = 1$.  In order to calculate the frequencies, we first need to determine the ground-state densities of bosons and fermions. As mentioned before, we accomplish this task by employing a variational approach similar to the one described in Ref. \cite{yabu1}. For fermions we apply TF approximation and within this approximation the density of fermions is given by
\begin{equation}
\rho_{f}({\bf r}) = \left (\frac{2m_{f}}{\hbar^{2}}\right )^{3/2}\frac{1}{6\pi^{2}}\left ( \mu_{f} - V_{f}({\bf r}) - g_{bf}\rho_{b}({\bf r})\right )^{3/2}.
\label{tfdensity}
\end{equation}
On the other hand, the density of bosons is determined by choosing appropriate variational forms for it. In particular for negative values of boson-fermion scattering length $a_{bf}$, we use single-parameter gaussian form for the boson density given by $\rho_{b}({\bf r}) = A_{b}e^{-\alpha r^{2}}$, where $A_{b}$ is the normalization constant fixed by the normalization condition $\int\rho_{b}({\bf r})d^{3}{\bf r} = N_{b}$ and $\alpha$ is the single variational parameter determined by minimization of the energy functional of Eq. (\ref{energyfunc}). It has already been shown \cite{tanatar} that gaussian ansatz leads to incorrect density profile for positive values of $a_{bf}$ specially when $a_{bf}\geq a_{bb}$. In order to circumvent this problem, we employ a two-parameter form
\begin{equation}
\rho_{b}({\bf r}) = A_{b}\left ( 1 - \frac{r^{2}}{R^{2}}\right )^{1 + \lambda}\theta (R - r)
\label{fetter}
\end{equation}
proposed by Fetter \cite{fetter} to study BEC. In Eq. (\ref{fetter}), $\theta (R - r)$ is the step function, $R$ and $\lambda$ are the two variational parameters again determined by minimization of the total energy. Here we mention that the ground-state calculations performed with this variational approach yields results which are quite close to the corresponding results reported in Refs. \cite{yabu2,roth} 

By using the ground-state densities obtained by variational approach we evaluate the required integrals to calculate the frequencies of the collective oscillations given by Eqs. (\ref{monopole}) and (\ref{quadrupole}). The results of these calculations are discussed in the next section.

\section{Results and discussion}
We begin this section with the discussion of results for the monopole mode of the collective oscillations followed by the results for quadrupole mode. Before discussing the results we note that in this paper all the numerical calculations are performed for $N_{b}=10^{6}$ and $N_{f}=10^{6}$. Following Ref. \cite{yabu2} the boson-boson scattering length is chosen to be $\tilde{a}_{bb} = a_{bb}/l_{b} = 0.2/(8\pi)$ ( where $l_{b} = \sqrt{\hbar/m_{b}\omega_{b}}$ ) and the boson-fermion scattering length is varied from a negative value characterizing collapse of the mixture to a large positive value at which Bose-Fermi mixture undergoes phase separation. Such wide variation of the boson-fermion interaction strength has recently been achieved by tuning the magnetic field at a Fesbach resonance \cite{zaccanti} We wish to emphasize here that the above choice of the parameters does not lead to any loss of generality of the results presented in this paper. 
  
\subsection{Monopole mode} In order to demonstrate the effect of different mass of Bose- and Fermi- particles, we first display the frequency and the mixing angle of the monopole mode as a function of coupling strength $\kappa = a_{bf}/a_{bb}$ in Fig. 1 for two different values of mass ratio, namely $m_{b}/m_{f}=1$ and $m_{b}/m_{f}= 87/40$. The latter ratio corresponds to the Bose-Fermi mixture of $^{87}Rb$ and  $^{40}K$ atoms. Fig. 1 clearly shows that the monopole mode of the collective oscillations exhibits significantly different behaviour when the ratio $m_{b}/m_{f}$ is changed from 1 to $87/40$.  The differences in the behaviour of the mixing angle which characterizes the nature of the mode of the collective oscillations are clearly visible even in the weak interaction regime ($\kappa<1$). For example, in the weak boson-fermion interaction regime the monopole mode is dominantly fermionic in nature for $m_{b}/m_{f}=1$, whereas it becomes dominantly bosonic for $m_{b}/m_{f}= 87/40$. Consequently, the frequency of the monopole mode shows completely different behaviour as the mass ratio is changed. This result has motivated us to investigate the effect of different mass of a Bose- and a Fermi-particle on the collective oscillations of  boson-fermion mixtures. At this stage it is natural to ask following question:  is it that any mass ratio different from unity will change the behaviour of monopole mode of the collective oscillations or there exists a critical value of mass ratio $(m_{b}/m_{f})_{c}$ at which change in the behaviour of monopole mode occurs? In order to answer this question, we focus our attention on the collective oscillations of non-interacting ($\tilde{a}_{bf} = 0$) boson-fermion mixture as we have already seen that effect of non-unity mass ratio is appreciable even for very small values of $\kappa$ around zero. Moreover, for non-interacting Bose-Fermi mixture one can derive analytical results for the frequencies and the mixing angle, which makes the analysis more lucid. By using the expression for the frequency of the monopole mode given by Eq. (\ref{monopole}) for non-interacting case the condition of extremum $d\omega_{m}/d\theta = 0$ leads to two possible solutions for $\theta$:  $\theta = \pm\pi/4$. Now to determine the value $\theta_{min}$ out of the above two values at which $\omega_{m}$ becomes minimum, we impose the condition  $\left (d^{2}\omega_{m}/d\theta^{2}\right )<0$ for $\theta_{min}$. This condition then leads to 
\begin{equation}
G_{mon}\sin 2\theta_{min} < 0,\qquad {\rm where} \qquad  G_{mon} = \left [ \frac{-T_{b}}{V_{b}} + \left ( 5 - 4\left (\frac{m_{b}}{m_{f}}\right )\right )\right ].
\label{monmin}
\end{equation}
From the above inequality, we infer that when $G_{mon}$ is positive (negative), the corresponding mixing angle of the monopole mode becomes $\theta_{min}=-\pi/4 (\pi/4)$. In the limit of TF approximation for bosons ($T_{b}/V_{b} = 0)$, Eq. (\ref{monmin}) reduces to
\begin{equation}
G_{mon}^{TF}\sin 2\theta_{min} < 0,\qquad {\rm where} \qquad G_{mon}^{TF} = \left [ 5 - 4\left (\frac{m_{b}}{m_{f}}\right )\right ].
\label{monmintf}
\end{equation}
For $m_{b}=m_{f}$, as considered in the previous works on the collective oscillations of Bose-Fermi mixtures \cite{yabu2,liu}, $G_{mon}^{TF}$ in Eq. (\ref{monmintf}) is positive and consequently the  mixing angle minimizing the monopole mode is $-\pi/4$. Therefore, the monopole mode for a non-interacting Bose-Fermi mixture with $m_{b}= m_{f}$ is purely fermionic in character and the frequency of the oscillation is given by $\omega_{m}=2\omega_{b}$. These results are clearly illustrated in Fig. 1 by the dashed curves. On the other hand, for $m_{b}\neq m_{f}$, it can be easily inferred from Eq. (\ref{monmintf}) that $\theta_{min}$ can assume value $-\pi/4$ or $\pi/4$ depending on the value of the ratio $m_{b}/m_{f}$. The transition from $\theta_{min}=-\pi/4$ to $\theta_{min}=\pi/4$ will take place at a critical value $(m_{b}/m_{f})_{c}=5/4=1.25$. For $m_{b}/m_{f}< 1.25$, $G_{mon}^{TF}$ is positive and consequently $\theta_{min}=-\pi/4$ indicating that the monopole mode is purely fermionic in nature similar to the case of $m_{b}=m_{f}$ . In contrast to this, for $m_{b}/m_{f} > 1.25$, the functin $G_{mon}^{TF}$ is negative and the Eq. (\ref{monmintf}) is satisfied for $\theta_{min}=\pi/4$ leading to a purely bosonic monopole mode with the frequency $\omega_{m}=\sqrt{5}\omega_{b}$. The results shown in Fig. 1 with solid curves correspond to this situation as $m_{b}/m_{f} = 87/40 > 1.25$.

When $T_{b}/V_{b} \neq 0$ the critical mass ratio is given by
\begin{equation}
\left (\frac{m_{b}}{m_{f}}\right )_{c} = \frac{5}{4} - \frac{T_{b}}{4V_{b}},
\label{moncritmass}
\end{equation}
and to determine the value of $\theta_{min}$ for general case, we plot $G_{mon}$ as a function of the mass ratio $m_{b}/m_{f}$ in Fig. 2. It can be seen from Fig.2 that even for the general case the value of the critical mass ratio is very close to TF case result $(m_{b}/m_{f})_{c} = 1.25$. Therefore, from the above results we infer that depending on the value of mass ratio $m_{b}/m_{f}$ the monopole mode of the boson-fermion mixture can be dominantly bosonic or fermionic in character. The transition from one to other occurs at critical mass ratio $(m_{b}/m_{f})_{c}=1.25$.

Having found the nature of the monopole mode for the non-interacting Bose-Fermi mixture, we next focus our attention on the interacting case ($\tilde{a}_{bf}\neq 0$). For the interacting case, we obtain  the value of $\theta_{min}$ by numerically minimizing the monopole mode frequency $\omega_{m}$. The results of these numerical calculations are also displayed in Fig. 1. We find that for large values of boson-fermion interaction strength both in the positive and the negative directions the monopole mode becomes coherent superposition of bosonic and fermionic oscillations. The proportion of mixing of these two types oscillations depends on the boson-fermion mass ratio and the interaction strength $\kappa$. The monopole modes corresponding to both $m_{b}/m_{f}<(m_{b}/m_{f})_{c}$ and $m_{b}/m_{f}>(m_{b}/m_{f})_{c}$ cases tend toward in-phase oscillations ($\theta_{min} = 0$) of the Bose-Fermi mixture as it approaches either the collapse or the spatial phase separation regimes. 
We note here that depending on the ratio of the Bose-Fermi mass the monopole mode follows either the lower branch characterized by purely fermionic oscillations at zero interaction to the coherent superposition of bosonic and fermionic oscillations  for finite values of interaction strength or the upper branch representing purely bosonic to the coherent superposition of bosonic and fermionic oscillations. Consequently, the frequency of the monopole mode show drastically different behaviours for mass ratios below and above the critical limit. Moreover, from the behaviour of the monopole mode discussed above, we note that for a given mass ratio the character of the monopole mode at zero boson-fermion interaction strength can be employed to predict the nature of this mode at finite values of the interaction strength. 
For example, another experimentally achieved Bose-Fermi mixture composed of $^{7}Li$ and $^{6}Li$ has mass ratio $m_{b}/m_{f} = 7/6$, which is less than 1.25. From the above discussion, we infer that at zero boson-fermion interaction strength monopole mode is purely fermionic in character and for finite values of interaction strength it will follow the lower branch similar to the case of $m_{b}/m_{f} = 1$ as shown by dashed lines in Fig. 1. 
\subsection{Quadrupole mode}
Now we consider the quadrupole mode of the collective oscillations of a trapped Bose-Fermi mixture and carry out an analysis similar to the one discussed above for the monopole mode. Using Eqs. (\ref{quadrupole}) and (\ref{xqyqzq}) for $\tilde{a}_{bf} = 0$, we find that for quadrupole mode also $\theta = -\pi/4$ and $\theta = \pi/4$ make the frequency $\omega_{q}$ extremum. The condition for obtaining mixing angle $\theta_{min}$ that minimizes the frequency of the quadrupole mode is given by
\begin{equation}
G_{quad}\sin2\theta_{min} < 0 \qquad {\rm where} \qquad G_{quad} = \left [ \frac{T_{b}}{V_{b}} + \left ( 1 - 2\left (\frac{m_{b}}{m_{f}}\right )\right )\right ].
\label{quadmin}
\end{equation} 
Within the TF approximation for the bosons above criterion reduces to 
\begin{equation}
G_{quad}^{TF}\sin2\theta_{min} < 0 \qquad {\rm where} \qquad G_{quad}^{TF} = \left [ 1 - 2\left (\frac{m_{b}}{m_{f}}\right )\right ]sin2\theta_{min}.
\label{quadmintf}
\end{equation}
For $m_{b}= m_{f}$ the function $G_{quad}^{TF}$ in Eq. (\ref{quadmintf}) is negative and the minimum condition is satisfied if $\theta_{min} = \pi/4$. Therefore, for non-interacting Bose-Fermi mixture the quadrupole mode is of purely bosonic character and the frequency of the oscillations is $\omega_{q} = \sqrt{2}\omega_{b}$ \cite{yabu2}. However, for  $m_{b}\neq m_{f}$, the quadrupole frequency can become minimum either for $\theta_{min} = \pi/4$ or $\theta_{min} = -\pi/4$ depending on whether $m_{b}/m_{f} > (m_{b}/m_{f})_{c}$ or $m_{b}/m_{f} < (m_{b}/m_{f})_{c}$ respectively, where $(m_{b}/m_{f})_{c}=0.5$. Thus, for $m_{b}/m_{f} < 0.5$, the quadrupole mode shows purely fermionic character and the corresponding frequency of the non-interacting  Bose-Fermi mixture is given by $\omega_{q} = 2\omega_{b}$. On the other hand, when $m_{b}/m_{f} > 0.5$, the quadrupole mode possesses purely bosonic character with frequency $\omega_{q} = \sqrt{2}\omega_{b}$. 

For the general case ($T_{b}/V_{b}\neq 0$) the critical mass ratio at which quadrupole undergoes a change in character is given by
\begin{equation}
\left (\frac{m_{b}}{m_{f}}\right )_{c} = \frac{1}{2} + \frac{T_{b}}{2V_{b}}.
\label{quadcritmass}
\end{equation}
In order to determine the mixing angle that minimizes the quadrupole frequency for the general case, we plot in Fig. 3 the $G_{quad}$ as a function of $m_{b}/m_{f}$. Again it can be seen that corresponding to $N_{b} = 10^{6}$ the critical mass ratio for quadrupole mode is very close the TF result $(m_{b}/m_{f})_{c}=0.5$. 

Like monopole mode, the mixing angle and the frequency of the quadrupole mode at finite value of the boson-fermion interaction strength are obtained numerically by minimizing the quadrupole frequency. The results of such calculation for two different values of mass ratio namely, $(m_{b}/m_{f})=1/3$ and $(m_{b}/m_{f})=1$ are shown in Fig. 4.  
These two figures clearly elucidate the results on the quadrupole mode  discussed above. For example, for $(m_{b}/m_{f})=1/3$ the minimum criterion Eq. (\ref{quadmin})is satisfied by $\theta_{min}=-\pi/4$ corresponding to a fermionic quadrupole mode with the frequency $\omega_{q} = 2\omega_{b}$ at zero value of the boson-fermion interaction.  For finite values of the boson-fermion interaction strength the quadrupole mode no longer remains a purely bosonic or fermionic in character, rather it becomes a coherent superposition of both. The proportion of the mixing two modes depends on the boson-fermion mass ratio and the interaction strength. Like monopole mode the superposed quadrupole mode either lie on the lower or the upper branch depending on the mass ratio at finite values of the boson-fermion interaction strength.
As a result of this the frequency of the quadrupole mode show different behaviour for $m_{b}/m_{f}<(m_{b}/m_{f})_{c}$ and $m_{b}/m_{f}>(m_{b}/m_{f})_{c}$ and this is clearly elucidated in Fig. 4.  

\section{Conclusion}
In this paper we have studied the effect of different mass of a Bose- and a Fermi-particle on collective oscillations of trapped Bose-Fermi mixtures. In particular we have considered the monopole and the quadrupole modes of collective oscillations and investigated the effect of unequal mass on the characters and the frequencies of the modes. The calculations of frequencies are performed by employing sum-rule approach of many-body response theory in conjunction with a variational method to obtain the ground-state densities of bosons and fermions. By considering the non-interacting Bose-Fermi mixtures we have determined the critical mass ratio both for the monopole and the quadrupole modes. At these critical values of $m_{b}/m_{f}$ the monopole and quadrupole modes undergo change in character and consequently the frequencies. In order to study the effect of unequal mass for the finite values of the boson-fermion interaction strength we determine the mixing angle characterizing the nature of the mode by numerical minimization of the corresponding frequencies. We have shown that the characters of both monopole the quadrupole modes of a Bose-Fermi mixture are decided by their corresponding nature at non-interacting regimes.  The recent achievement of trapped degenerate Bose-Fermi mixtures of atomic gases in laboratory and the capability of tuning the boson-fermion interaction strength by Fesbach resonance make the experimental verification of the results presented in this paper feasible.

\acknowledgments{ I wish to thank Dr. S. C. Mehendale and Dr. M. P. Singh for critical reading of the manuscripts and also for several useful discussions}  


\clearpage
\newpage
\section*{Figure captions}
{\bf Fig.1} The frequency  (upper part) and the mixing angle (lower part) of the monopole mode as a function of the boson-fermion interaction strength $\kappa$ for two different values of $m_{b}/m_{f}$ : $m_{b}/m_{f} = 1$ ( dashed curve) and $m_{b}/m_{f} = 87/40$ ( solid curve). The horizontal upper and the lower dotted lines in the bottom panel denote the mixing angles for pure bosonic and fermionic modes respectively. 

{\bf Fig.2}Plot of function $G_{mon}$ of Eq. (\ref{monmin}) as function of the ratio $m_{b}/m_{f}$. The intersection with the dotted horizontal line denotes the critical value of ratio of masses of a boson and a fermion $\left(m_{b}/m_{f}\right)_{c}$.

{\bf Fig.3}Plot of function $G_{quad}$ of Eq. (\ref{quadmin}) as function of the ratio $m_{b}/m_{f}$. The intersection with the dotted horizontal line denotes the critical value of ratio of masses of a boson and a fermion $\left(m_{b}/m_{f}\right)_{c}$.
  
{\bf Fig.4} The frequency  (upper part) and the mixing angle (lower part) of the quadrupole mode as a function of the boson-fermion interaction strength $\kappa$ for two different values of $m_{b}/m_{f}$: $m_{b}/m_{f} = 1$ ( dashed curve) and $m_{b}/m_{f} = 1/3$ ( solid curve). The horizontal upper and the lower dotted lines in the bottom panel denote the mixing angles for pure bosonic and fermionic modes respectively.   
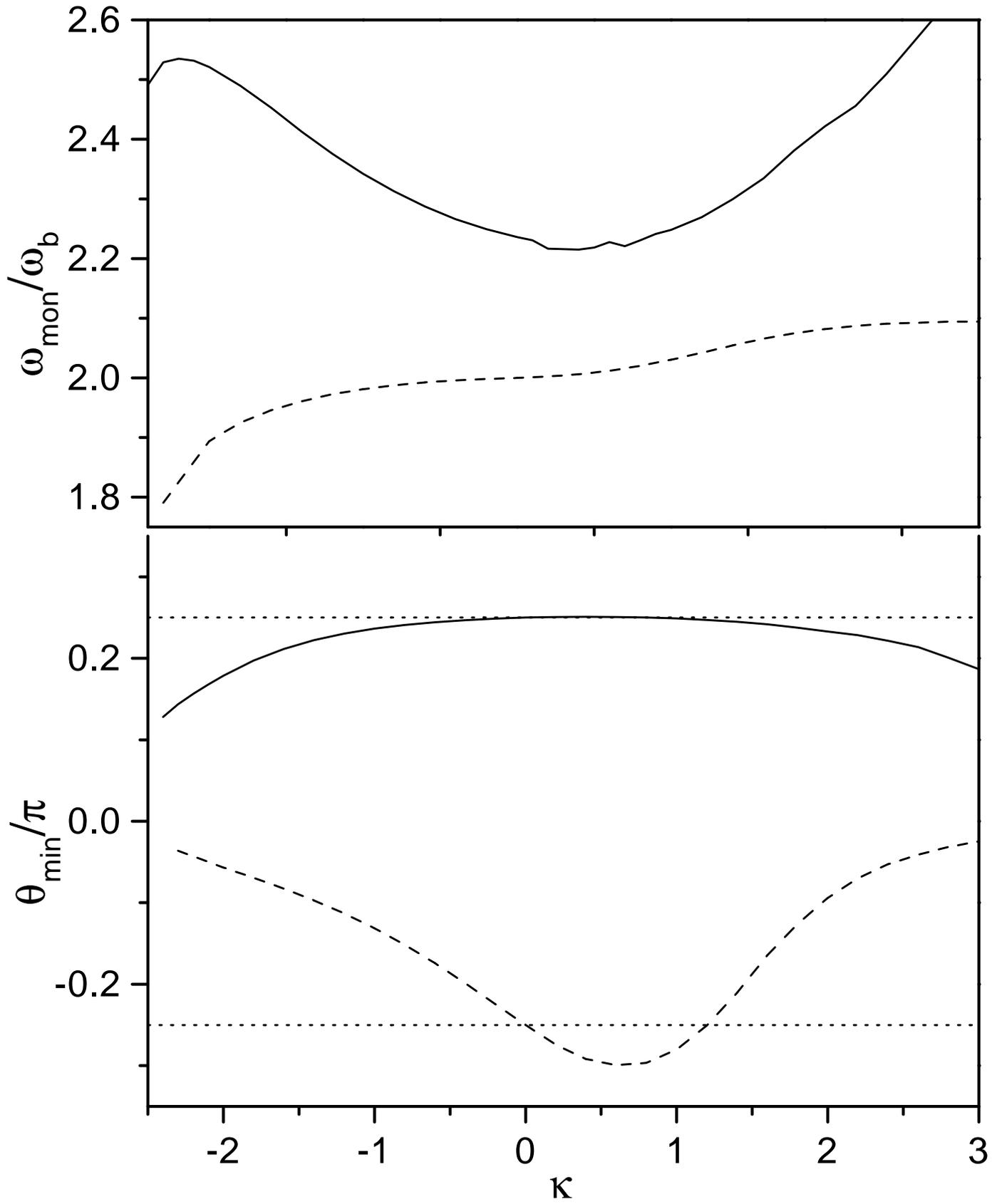
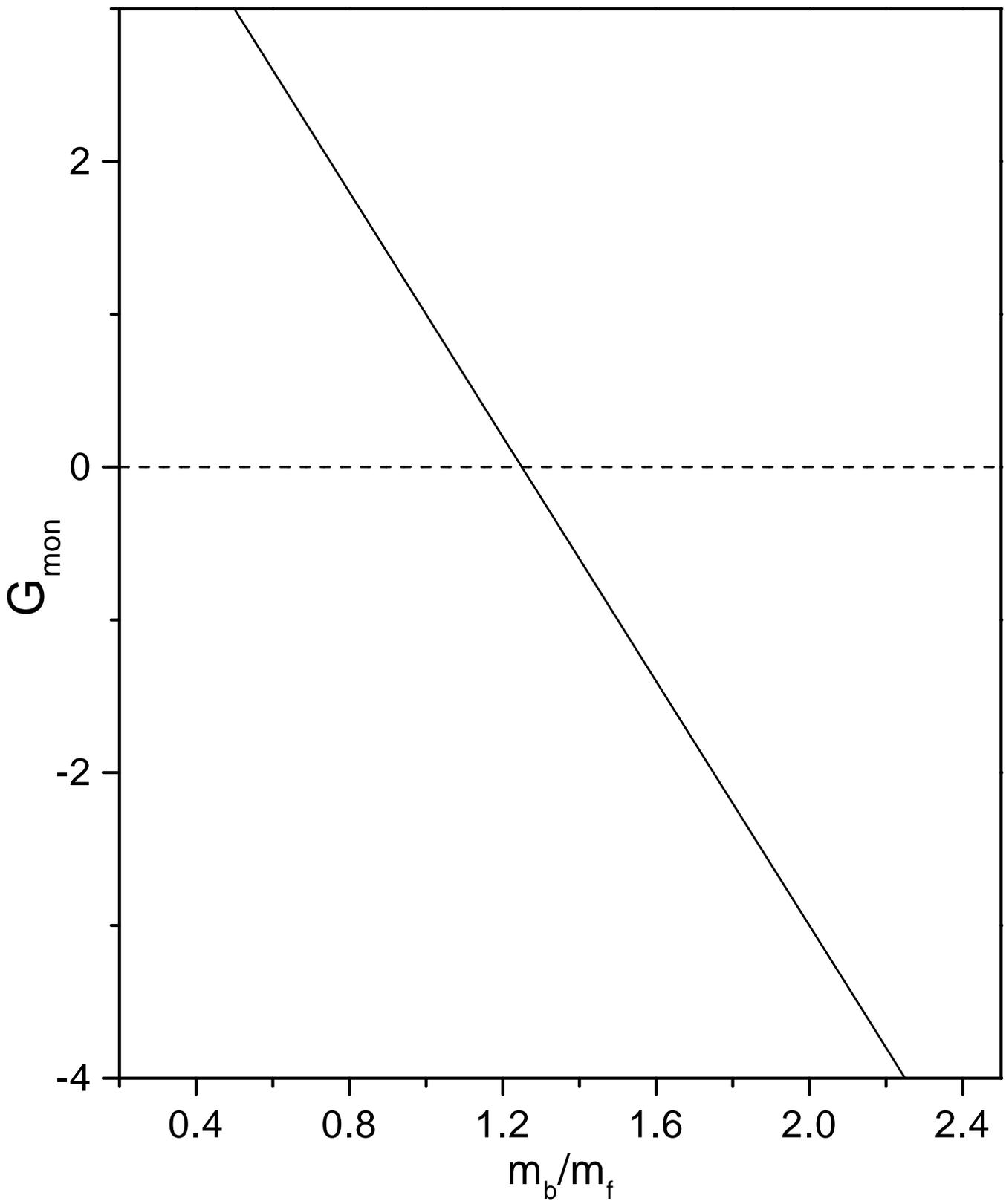
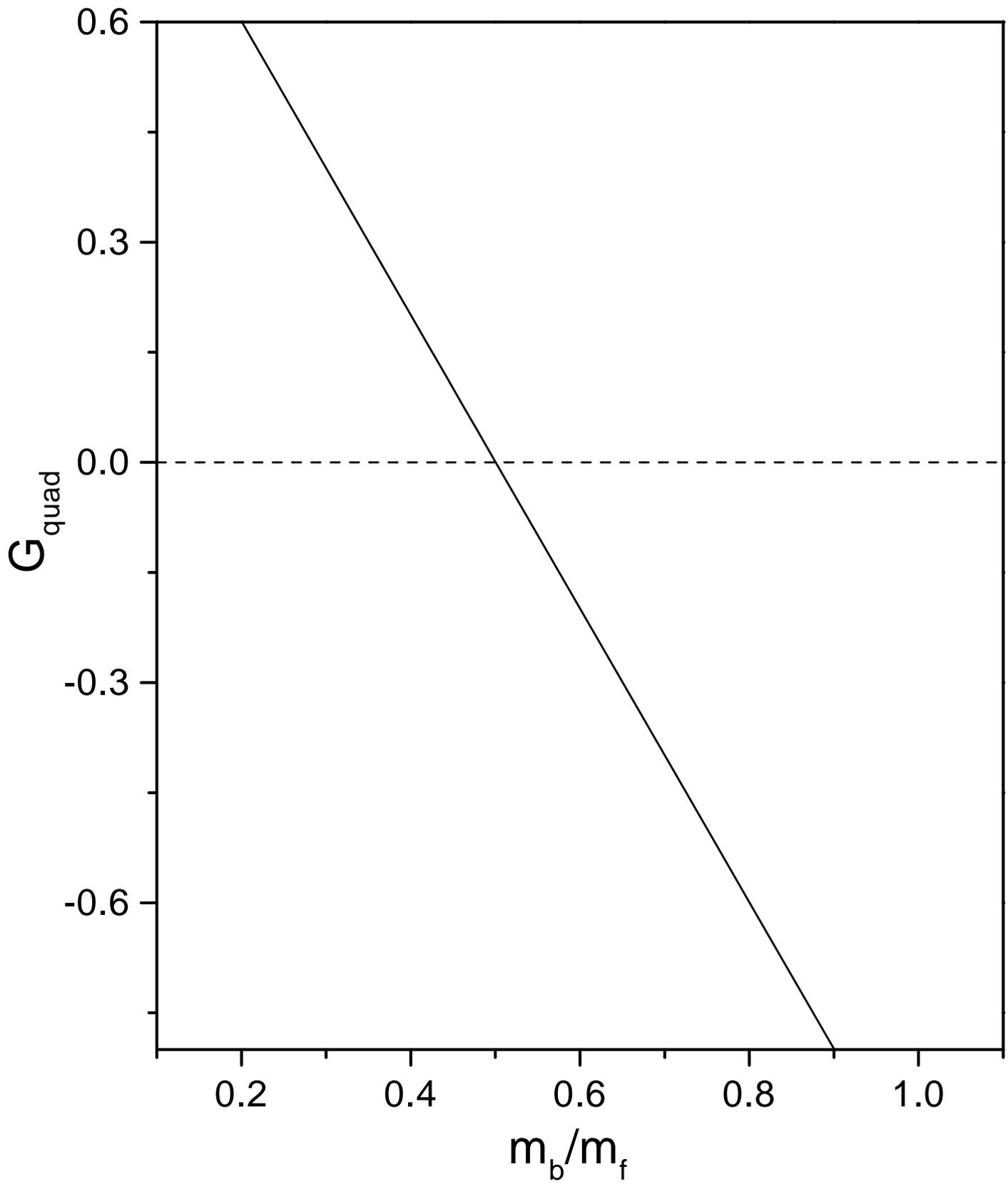
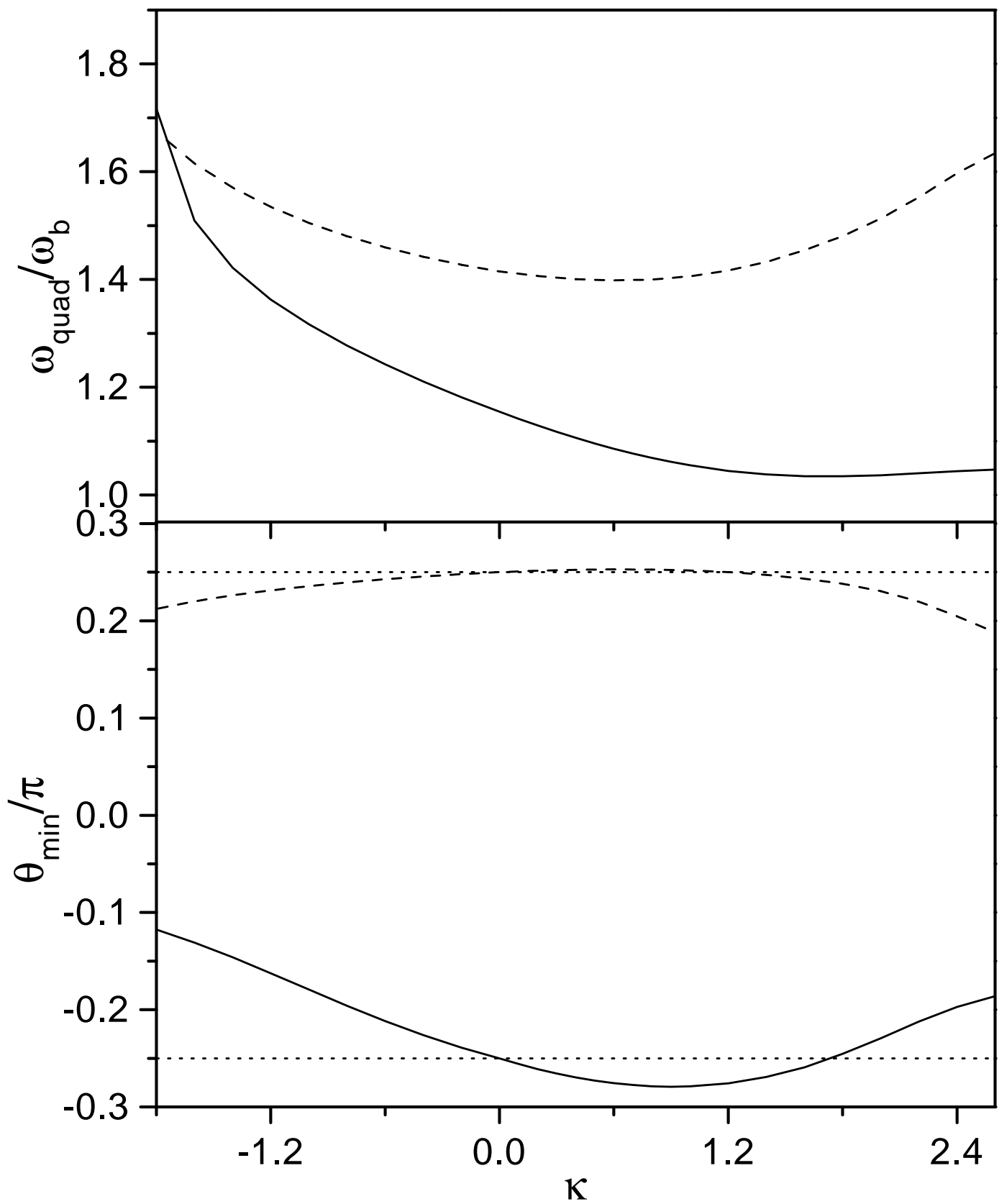
\end{document}